\documentclass[11pt]{article}

\usepackage[english]{babel}

\usepackage[a4paper,top=2cm,bottom=2cm,left=2cm,right=2cm,marginparwidth=2.25cm]{geometry}

\usepackage{amssymb}
\usepackage{siunitx}
\PassOptionsToPackage{hyphens}{url}\usepackage{hyperref}
\usepackage{cleveref}
\usepackage[utf8]{inputenc}
\usepackage{csquotes}
\usepackage{booktabs}
\usepackage{longtable}
\usepackage{adjustbox}
\usepackage{array}
\usepackage{url}
\usepackage{titlesec}
\usepackage{authblk}
\usepackage{microtype}
\PassOptionsToPackage{hyphens}{url}
\usepackage{hyperref}

\usepackage[normalem]{ulem}

\usepackage[english]{babel}

\title{Building a continuous benchmarking ecosystem in bioinformatics}

\author[1,2,*]{Izaskun Mallona}
\author[2,3]{Charlotte Soneson}
\author[1,2]{Ben Carrillo}
\author[1,2,4]{Almut Lütge}
\author[1]{Daniel Incicau}
\author[1,2]{Reto Gerber}
\author[1,2]{Anthony Sonrel}
\author[1,2,*]{Mark D. Robinson}

\affil[1]{Department of Molecular Life Sciences, University of Zurich, 8057 Zurich, Switzerland}
\affil[2]{SIB Swiss Institute of Bioinformatics, University of Zurich, 8057 Zurich, Switzerland}
\affil[3]{Friedrich Miescher Institute for Biomedical Research, 4056 Basel, Switzerland}
\affil[4]{Swiss Data Science Centre, 8092 Zurich, Switzerland}

\affil[*]{Correspondence: izaskun.mallona@gmail.com,mark.robinson@mls.uzh.ch}

\begin{document}
\maketitle


\begin{abstract}
Benchmarking, which involves collecting reference datasets and demonstrating method performances, is a requirement for the development of new computational tools, but also becomes a domain of its own to achieve neutral comparisons of methods. Although a lot has been written about how to design and conduct benchmark studies, this Perspective sheds light on a wish list for a computational platform to orchestrate benchmark studies. We discuss various ideas for organizing reproducible software environments, formally defining benchmarks, orchestrating standardized workflows, and how they interface with computing infrastructure.
\end{abstract}


\pagebreak

\section*{Introduction}
\label{sec:introduction}

Benchmarking (defined below) is a critical step in method development. Such studies are typically published as either `methods-development papers' (MDPs) where new methods are compared against existing ones; or `benchmark-only papers' (BOPs), where a set of existing methods are compared in a more neutral way, as defined in \cite{cao2023-jz}. A benchmarking study can be run by a single person  (solo benchmarking) or a small group, or it can be organized as a `challenge' or within a hackathon. In this work, we present a perspective on building a robust benchmarking ecosystem in bioinformatics, coming primarily from the methodologist's viewpoint (researchers developing computational tools).  While our focus is bioinformatics, we aim to abstract principles and connect to similar challenges in adjacent fields. 

In our discussion, we outline concepts related to workflow automation, benchmark formalization and benchmarking-specific tasks beyond the former, including philosophical issues like gatekeeping and trust; and provide some technical recommendations, such as standardization.

\section*{Why a benchmarking system?}

\begin{figure}
    \centering
    \includegraphics[width=\linewidth]{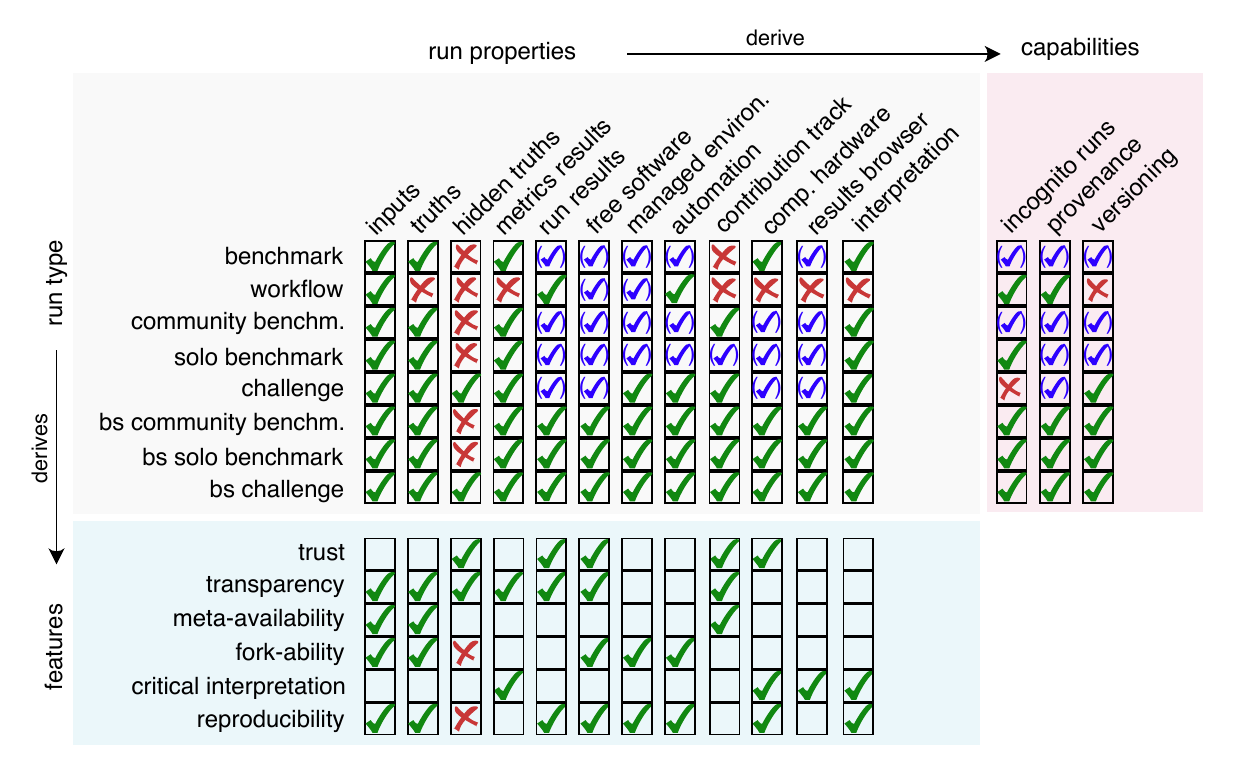}
     \caption[Run types, properties, capabilities and features.]{Solo and community benchmarks, workflows, and benchmarking-system-enabled (\textit{bs}) benchmarks all differ in run properties and their consequences in terms of capabilities and features, including whether: inputs, truths, hidden truths, metrics or run results are available for readers; code is FOSS, software environments are managed, and execution automated; there is a clear track of (human) contributors; results are run in comparable hardware; an entrypoint to results or result interpretation are provided. These run-intrinsic properties have consequences in terms of features (i.e., transparency, forkability, availability for metaanalysis) and technical capabilities (i.e. software versioning or ability to run locally without leaving traces - in \textit{incognito} mode). Green, blue and red ticks depict available, perhaps available and unavailable features, respectively. \label{fig:definitions}}
    
\end{figure}

We define a benchmark as a conceptual framework to evaluate the performance of computational methods for a given task, though it can extend to assessing other aspects, like data-generating technologies. A benchmark thus requires a well-defined task and typically a definition of correctness, or ground-truth, to be precisely defined in advance  (Figure~\ref{fig:definitions}). A benchmark then consists of benchmark `components', such as simulations to generate datasets, preprocessing steps, methods, and metrics. 

\begin{figure}
    \centering
    \includegraphics[width=0.8\linewidth]{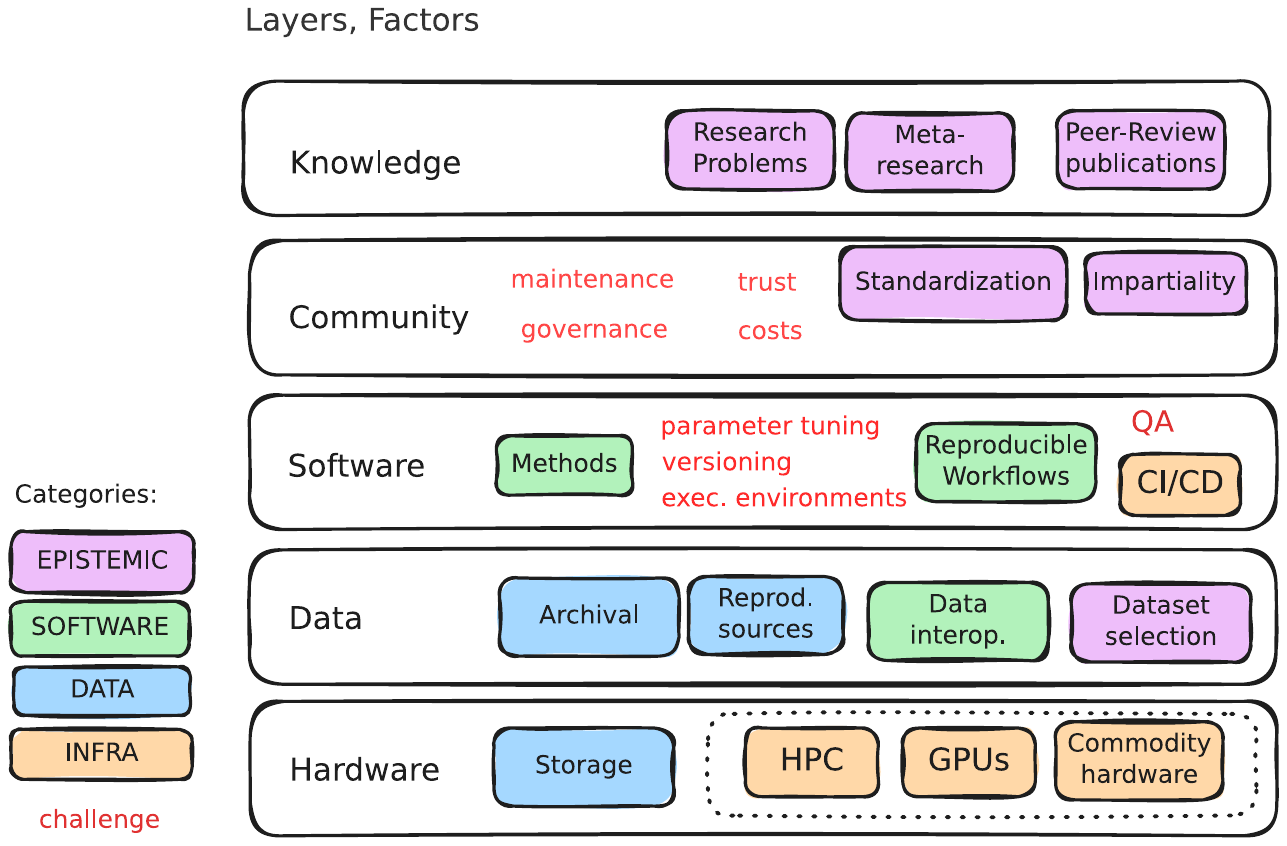}
    \caption[Benchmarking conceptual layers and their challenges.]{Benchmarking conceptual layers and their challenges. The process of benchmarking is multilayered, with layer-specific factors belonging to different categories (colored boxes) and challenges (in red). \emph{Hardware}: infrastructure and its costs. \emph{Data}: dataset archival, openness, interoperability, and selection.  \emph{Software}: method implementations, reproducibility, workflow execution, continuous integration and delivery (CI/CD), versioning, quality assurance (QA). \emph{Community}: Standardization and impartiality (governance, transparency, building trust, long term maintainability). \emph{Knowledge}: research and meta-research, academic publications.}
    \label{fig:layers}
\end{figure}

Can benchmarking be systematized? Benchmarking aims to calculate, collect and report performance metrics of methods aiming to solve a task. Such a process normally requires collecting appropriate data and running methods in an automated manner. Some of these tasks can be automated by workflow systems. Others, such as sharing results in an interactive manner and, particularly, result interpretation and making results reusable by others, go beyond what workflows offer. Similarly, collaborative benchmarking and challenges add extra layers of complexity, including collaboration tracking, hiding ground truths, or the ability to release multiple versions of the same benchmark as it develops (Figure~\ref{fig:definitions}). Moreover, scientific benchmarking borrows procedures from the scientific method and from scientific publishing, hence having expectations of fairness, reproducibility, transparency, and trust. \emph{Benchmarking systems} aim to orchestrate the workflow management, benchmarking and community engagement aspects of the process to generate benchmark `artifacts' (code snapshots, file outputs to be shared, performance outputs) in a systematic way and following good practices and standards. Namely,  the various elements of code, especially the details on how datasets are preprocessed and how methods are run, should be explicitly available for scrutiny. Such systems should facilitate components to be public, open and allow contributions (e.g., corrections or new components), such that the benchmarks of the future can run more continuously. At the outset of a benchmark, a benchmark `definition' could be envisioned, which could give a formal specification of the entire set of components (and the pattern of artifacts to be generated). The benchmark definition could even be expressed as a single configuration file that specifies the scope and topology of components to include, details of the repositories (code implementations with versions), the instructions to create software environments (accessible across compute architectures), the parameters used, and what components to snapshot for a release. A benchmark definition in terms of layers and the corresponding challenges and opportunities within each layer is illustrated in Figure~\ref{fig:layers}.  

\section*{Benchmark stakeholders}

Benchmarks are useful for a broad set of stakeholders. For a \emph{data analyst}, with an aim to perform a specific type of analysis on a specific dataset, benchmarks are helpful tools to find a suitable method for the task. Hence, it is important that benchmarks include datasets similar to the one of interest to the analyst, as method performance often depends on characteristics of the data. Thus, a thorough characterization of the datasets considered in a benchmark is desirable. Given that benchmarks typically evaluate methods using a range of metrics and datasets, not all of which are relevant to all analysts, flexibility in defining the ranking approach and aggregation of metrics is preferable. Such flexibility is typically not part of published benchmarks today \cite{Sonrel2023-te}. However, a well-structured benchmarking system could accommodate flexible filtering and aggregation of metrics and datasets, and additionally provide access to the code and software stack that the analyst would need to apply the selected method to their own dataset. 

For \emph{method developers}, benchmarking allows their method to be compared to the state of the art, ideally using neutral datasets and metrics to avoid intrinsic bias \cite{Boulesteix2013-vy, Weber2019-el}; a recent call for reporting checklists was also made \cite{Brooks2024-rq}. A diverse, accessible set of benchmarking datasets and metrics can address this bias and lower the entry barrier for method developers. Moreover, it is important that the new method is compared to the current state of competing methods, to best mimic the choices available to a potential method user. Alternatively, new methods could be compared to previous versions or even different implementations of the `same' method \cite{Rich2024-jh}, to illustrate stability of code changes or improvements made over time. In both cases, the typical approach today would be for the method developer to rerun all the methods and versions to which they want to compare their current method, on all the datasets they wish to use. This can be challenging and time consuming, since not all methods run on the same hardware setup or in the same software environment (e.g., R and Python versions, library dependencies). It also leads to a lot of redundancy across method comparisons (multiple stakeholders implementing similar workflows), which could be reduced by the presence of a benchmarking system where such results would be accessible and extendable. Finally, the results of the benchmark of a new method should be included in the corresponding publication, and thus an easy way of including the new method into the set of results served by the benchmarking system, and generating a snapshot in time for reproducibility, would be of great help to the developer. 

Benchmarks can also be of high value to \emph{scientific journals} and \emph{funding agencies}. Well-executed benchmark studies can be highly utilized and influential, and may point to gaps in the current body of methods, guiding future methodological developments; at the same time, in fast-moving fields like bioinformatics, benchmarks have the tendency to rapidly become stale. Journals and funding bodies also have a vested interest in ensuring that published or funded method developments are performed to a high standard, that unnecessary redundancy is reduced, and that results are FAIR (findable, accessible, interoperable and reusable) for maximal benefit to the community \cite{Wilkinson2016-bh}. Recently published benchmarks and challenges have been conceived as or ported to benchmarking systems to ensure these properties \cite{Nevers_2022,Bryce_Smith_2023, Pardo_Palacios_2023}. Thus, quality assurance, neutrality and transparency of benchmarking systems become highly relevant. 

Collaborative benchmarking efforts enunciate field-specific challenges and provide a venue for \emph{communities} to address them. For instance, DREAM \cite{meyer2021advances}, GIAB \cite{zook2019open} and CASP \cite{moult2005-ne} challenges and benchmarks have demonstrated how collaborative benchmarks can help in moving fields forward.

Finally, benchmarks could form a research topic on their own, either for newcomers to the field to get acquainted with the conceptual options available, or for experts to consolidate research efforts of a subfield. For this, it is important that there are suitable venues for publishing benchmarks. Regardless, a \emph{benchmarker} (i.e., researcher leading a benchmarking study) would also benefit from using a benchmarking system and curated collections of benchmark artifacts. Such benchmarkers may be good candidates for curating and maintaining such collections by designing and leading benchmark efforts as well as guiding other contributors to adhere to a high standard. 

\section*{Benchmarks can be formally defined}

Benchmarks are collections of data and source code, ideally executed as workflows within a computing environment. Indeed, recent studies highlight that code is frequently open, reusable and versioned, as are input datasets \cite{cao2023-jz, Sonrel2023-te}. However, \textit{extensibility} of benchmarking studies is generally very low, as is the proportion of benchmarks using a formal workflow system. 

\begin{figure}
    \centering
    \includegraphics[width=0.75\linewidth]{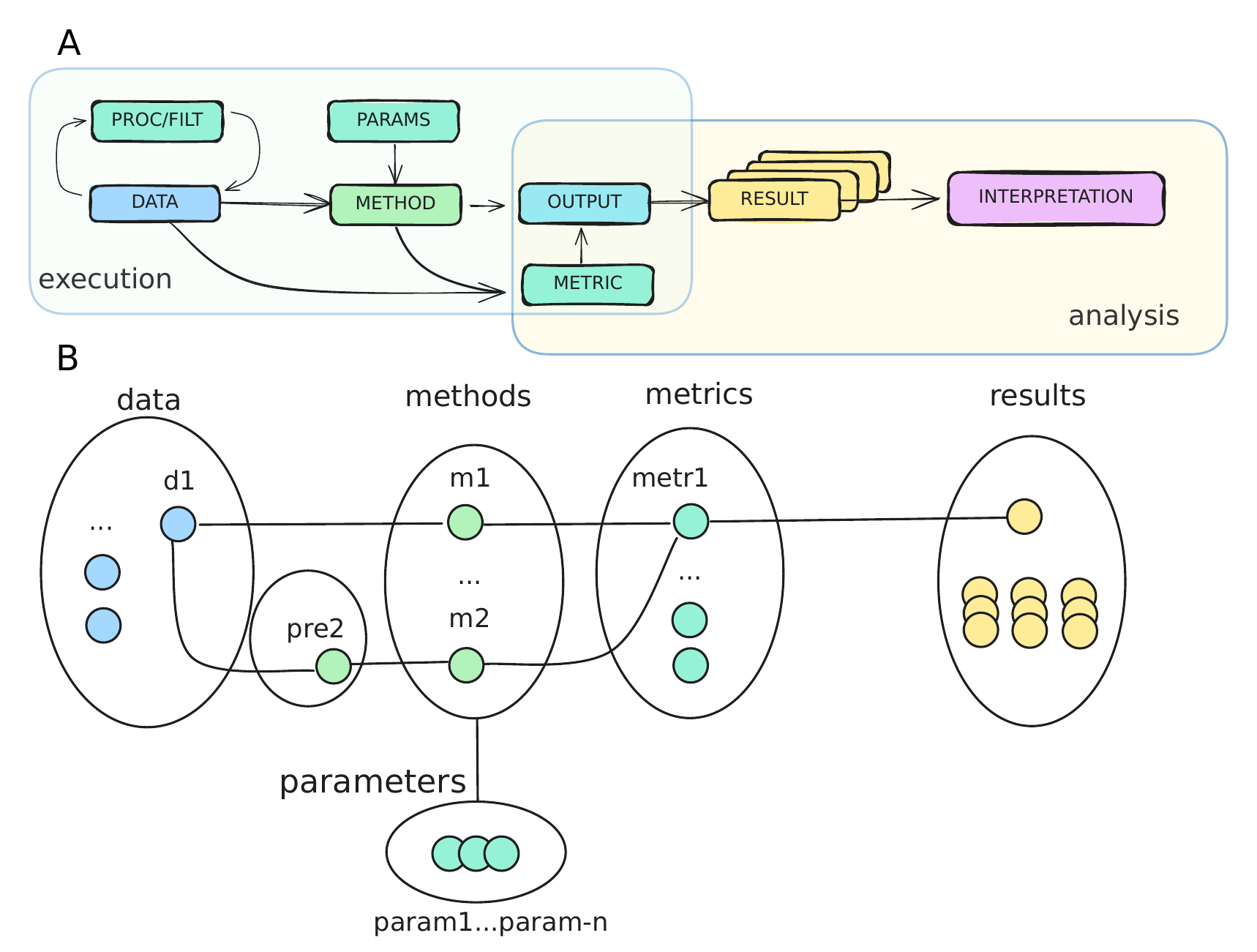} 
    \caption[Benchmark formalization.]{Benchmark formalization. \textbf{A}) Running a benchmark requires running a workflow (execution phase) to generate results to be critically evaluated (analysis phase). \textbf{B}) the execution phase consists on a mapping of methods to specific input files to generate output files, with optional steps to harmonize data and parametrize runs.}
    \label{fig:formalization}
\end{figure}

Over 350 workflow languages, platforms or systems  \cite{Wratten2021-el, Amstutz2024-qk} and one standard (Common Workflow Language, CWL) exist \cite{Amstutz2016-vo}. Some workflow languages can export workflow definitions to CWL, hence helping computational FAIR guidelines, i.e., bundling and processing data together with their metadata, tracking new metadata generation, recording provenance, and being accessible \cite{Goble2020-ps}. For benchmarking, we believe both the workflow layout and provenance can be formally defined (Figure~\ref{fig:formalization}).

Benchmarking as a process deals with tasks other than workflow definition and execution, including: collecting and keeping track of contributions from users; provisioning appropriate hardware; handling a reproducible and efficient software stack (the advantages of decoupling environment handling from workflow execution are further discussed in the section~`Reproducible software environments'); managing storage and access-control lists; rendering results dashboards and summarizing results; versioning code, workflow runs, and files; and providing extensive documentation from defining scope or contribution guidelines, to transparent logging and crediting contributions and disseminating results (Figure~\ref{fig:definitions}).

Currently, most benchmarking systems aim to enhance the workflow formalization and execution with strategies to fulfill benchmarking-specific needs. To name a few of these systems, \texttt{ncbench} bundles workflow specification and software management with \texttt{Snakemake} to benchmark result visualization with \texttt{Datavzrd} \cite{Hanssen2023-uf}. \texttt{OpenEBench} runs workflows in any workflow language, frequently \texttt{nextflow}, and makes use of \texttt{openVRE} as GUI and means of visualization \cite{Capella-Gutierrez2017-dh}. \texttt{OpenProblems} uses \texttt{Viash} \cite{cannoodt2024viash} and \texttt{nextflow} to handle software environments and workflows, respectively, and reports updated leaderboards with custom code \cite{luecken2025open}. To our knowledge, no benchmarking platform makes use of a fully expressive benchmark definition formalization language covering workflow generation and benchmarking-specific tasks. Orthogonally, some components of the benchmarking process have been formalized: \texttt{Viash} \cite{cannoodt2024viash} systematizes workflow components; \texttt{PROV-O} represents provenance data \cite{LeboUnknown-om}; and \texttt{Research Objects} (\texttt{RO} \cite{bechhofer2013linked}, e.g., \texttt{RO-Crate} \cite{soiland2022packaging}) specify a system to store data (files) and their metadata. Reusing pre-existing formalizations and extending them to the benchmarking field would facilitate conceptualization, community gathering and result sharing.

\section*{Scope and interpretation of benchmarking results}

Benchmarks are often focused on finding the `top' performing method(s), or on generating rankings. While this is valid in a few fields with clearly defined goals, such as image recognition \cite{Khan2018-ot} or protein structure prediction \cite{Jumper2021-oq}, bioinformatics tasks are typically evaluated by multiple metrics, some of which are not well understood \cite{Lutge2021-mt, Reinke2024-dw}. Additionally, ground truths are not always well-defined (e.g., derived with uncertainty from other experimental data) and multiple tasks (e.g., represented by different reference datasets) may be evaluated for a single method and several ways to assess performance are possible. Even when accompanied by dataset- and/or metric-specific performance measures, scaling and aggregation of results make it difficult to unpack fine-grained nuances of performance from a final summary table reported as a FunkyHeatmap \cite{funkyheatmap} as in most benchmarks (e.g., \cite{Saelens2019-jy}). Strobl \textit{et al.} push back against the `one method fits all datasets’ philosophy and advocate for highlighting dataset or parameter properties that are predictive of method performance \cite{Jelizarow2010-kr, Boulesteix2010-zj, Strobl2024-lp}. This aligns with what a data analyst wants: to identify the best performers in settings most similar to their problem at hand. This requires not only a thorough benchmark design, e.g., tracking of meaningful dataset metadata \cite{Strobl2024-lp}, but also can be supported by a benchmarking system that facilitates flexible navigation of results. 

Besides interactive dashboards for more nuanced result exploration \cite{bettr}, multi-criteria decision analysis (MCDA) \cite{Taherdoost2023-wd} has been proposed to guide users through benchmark results and multidimensional scaling (MDS) has been used to differentiate the effect of single datasets or metrics and recognize patterns beyond aggregated performance \cite{Niessl2022-fk}. An extensive analysis and presentation of the results helps not only users in the long term but also the whole scientific field to identify limitations across various use cases. The first advantage of such a strategy is to force the user to an attentive reading of the performance results so that the correct use case and relevant methods can be identified. The second, more important, advantage of presenting benchmark results extensively is to identify the nuances, e.g., the scenarios, types of datasets, or aspects of performance that are systematically associated with poor results. Being public and open to contributions from specialists of different fields, a benchmarking system could gather more reference datasets and methods, and allow the current state of the field to be better grasped. 

Parameter choices influence results interpretation, as they can be included in a benchmark to test performance across a wider range of settings, or to find the optimal parameter settings. 
Dataset-specific parameters, e.g., the true number of clusters when evaluating clustering algorithms, are not method-specific; they can be seen as (dataset-specific) metadata. To differentiate between the effect of the parameter and the methods performances themselves, data-specific parameters should be shared across methods during evaluation.
Similarly, optimizing method-specific parameters for some but not other methods can skew results and should be avoided. 
Aside of optimal and method-specific parameter searches, comparing a method under different settings against other methods can be viewed as comparing fully distinct methods, which should be clearly communicated in the results. 
For continuous and collaborative benchmarking, parameter optimization and usage should be governed and coordinated between all contributors, ideally ensuring comparability and interpretability of the results. 

Aside of task-tailored scores (i.e., intrinsic or extrinsic cluster validity measures for clustering tasks), collecting metrics to report computing performance (i.e., wallclock, on-disk read-write operations), requirements (e.g., GPUs, minimum memory) and their scalability depending on the input size (e.g., whether runtimes increase linearly, quadratically, etc) is key during benchmarking. To measure these performance metrics accurately, benchmarking systems ideally would provide means to: i) characterize the hardware running the benchmark, so method runtimes are comparable;  ii) track performance (CPU or GPU usage, memory) when running each benchmark component; iii) introduce minimal overhead during profiling; iv) collect variability in performance by running each method several times; and, v) go beyond reporting average resource usage and provide them over execution time (e.g., so embarrassingly-parallel algorithms can be differentiated from largely single-cored runs with few bursts of parallel execution).

\section*{Collaboration, gatekeeping and benchmarking infrastructure go hand in hand}

In designing a benchmarking system, it may be helpful to specify whether the benchmark is intended to be run `centralized' or `decentralized'. These terms can take different meanings depending on the context, but here the benchmark infrastructure location is intended. In most systems, code will be tracked in a \texttt{git} repository (e.g., on GitHub, GitLab). The workflow execution and the storage of derived files are typically located either in local deployments (e.g., \texttt{OEB-VRE} for \texttt{OpenEBench}), commercial clouds (e.g., AWS for \texttt{OpenProblems}), commercial continuous integration and continuous delivery (e.g., GitHub actions for \texttt{ncbench}) or on arbitrary infrastructure (i.e., laptop, server or cloud; e.g., omnibenchmark \cite{omnibenchmark}). 

Traditionally, benchmarks (whether BOP or MDP) are conducted in a decentralized fashion and typically by a small number of researchers (e.g., for MDP, the new method developer). This brings some disadvantages. For example, decentralization can be hugely inefficient; researchers studying the same task would collect reference datasets and code snippets to run methods, and implement a bespoke strategy (e.g., shell scripts) to orchestrate the methods run on datasets. Working together would mean sharing not only datasets and code snippets, but also execution environments (software but perhaps also hardware), alignment on parameters to evaluate and eventually sharing, consolidating, and interpreting results. However, because there are not many standards of how benchmarks are implemented \cite{Sonrel2023-te}, another disadvantage of decentralized benchmarks is that, while these components may eventually be openly shared as a product of a research study, the contents of a benchmark are non-interoperable, even across benchmarks of the same task. Furthermore, decentralized benchmarks have been criticized on the basis of `inflating model performance' due to decisions made in preprocessing or parameter tuning \cite{luecken2025open}; such issues could be mitigated by gathering the `wisdom of crowds' that is implicit in running centralized benchmarks.

Most benchmarks could begin decentralized, allowing for concept definition and preliminary results before `going public' (centralization phase). Users need a system that supports building benchmarks both ways, transitioning to centralization as the design matures. However, becoming centralized is not cost-free: it implies gatekeeping (contributors authenticate to the system and get access to computational resources) and requires additional measures that build trust (e.g., transparency).

Compute-intensive benchmarks are often run using cloud or high performance computing, further extending the challenge to ensure hardware homogeneity (for comparable performances) and adding complexity to the benchmarking system capabilities to automate node provisioning (or node selection) and to group executions such that costs are minimized. Some workflow managers, such as \texttt{Snakemake}, provide plugins to execute on the cloud or on-premises queuing systems, effectively decoupling workflow specification and execution.

\section*{Building trust and bringing together communities}

An active community is vital to the success of a benchmarking system. Benchmarkers (which may include multiple scientists) are responsible for planning and coordinating a benchmark, defining the task, possibly splitting it into subtasks or processing stages, and defining the data formats across the stages; the benchmarker also brings an authority role of how to review and approve contributions. Contributors (which may include method developers) curate and add content to the benchmark, which could be new datasets, methods or metrics, adhering to the guidelines set up by the benchmarker. Finally, the viewers (which may include benchmarkers and contributors) of benchmark results are users who retrieve one or more of the artifacts (code snippets, dataset files, intermediate results, or metric scores). These could span a range of use cases, including the data analyst choosing which method to use for a specific application, an instructor retrieving a curated dataset for teaching purposes, or a methods researcher prototyping their new method.

For assembling and sustaining communities, strategies include organizing seminars or workshops (on the educational side) or  hackathons and challenges (aimed at practical onboarding or contributions). Career-related or financial incentives, such as the prospect of a scientific publication, or compensation for the allocated time, are also common motivators. For a method developer, contributing to a public benchmark provides a way to test their method without the need to set up all the infrastructure themselves, and also provides an opportunity to advertise their work. Ultimately, the likelihood that someone will contribute to a benchmark will be associated with whether the results are of interest to them, as well as how easy it is to contribute. Contributions can be made easier by providing good organization, suggested `good first issues' to tackle, contributor guidelines and a code of conduct.

Authorship, e.g., acknowledging contributions, can be captured, reported and credited using third-party platforms such as Apicuron \cite{hatos2021apicuron} or by bundling standardized contribution metadata such as the Citation File Format (\texttt{CFF}) \cite{druskat_2021_5171937}. In this way, individual contributions to a benchmark can be acknowledged, browsed, and ideally propagated to academic deliverables (e.g., datasets, publications).

Publicly soliciting contributions poses challenges, requiring a transparent setup to build trust. Benchmarkers must trace and test contributions to detect misuse and prevent non-functional or suboptimal implementations. This can be addressed by a `quarantine zone', where new content is automatically tested before being added to the system. Similarly, a contributor should be able to trace their contribution, to ensure that no unintended modifications are made after submission. Ultimately, many risks can be mitigated by transparency \cite{Greenstein2016-os}, and feedback loops with (all) developers.

A key challenge is selecting datasets and metrics for a benchmark. Diversity of these ensures generalizability and prevents overfitting, but low-quality or redundant datasets can lead to misleading results. Possible mitigations include sensitivity analyses, evaluating the impact of single datasets or metrics on overall performance metrics, or reducing redundancy by down-weighting similar metrics to avoid a specific aspect dominating the evaluation. From a wider perspective, these questions are all related to the governance of a benchmark, a topic that has not received much attention. For governance, benchmarkers could choose to have the ultimate authority to decide on what gets included in a benchmark, simplifying the decision making but creating a risk of gatekeeping; or, decisions could be made collectively by a council of contributors, which may mitigate the risk of gatekeeping at least partially, but creates a more complex decision structure and introduces the need for a conflict resolution strategy. The governance model of a benchmark could also cover aspects related to computing infrastructure and storage, as well as guidelines to manage authorship of any resulting publications or resources. Having the governance model spelled out early in a project is likely to reduce friction later.

Neutrality is a highly desirable property in a benchmark: it means that efforts have been taken to ensure impartiality or unbiasedness (or at least transparency) of the various choices made (e.g., what datasets to include, how exactly to run methods, how to evaluate methods) \cite{Weber2019-el, Jelizarow2010-kr}. One could claim that all MDP benchmarks are non-neutral, since the goal is generally to highlight the virtues of a new method. Similarly, ground truth generation and simulations in MDP benchmarks can unconsciously mirror the method capabilities, mechanisms, or strengths, thus biasing evaluation. Furthermore, there are many field experts who may wish to conduct a benchmark (or participate in a collaborative benchmark) on analyses that involve their own tools, where neutrality is then harder to establish. Here, recent work to mitigate this conflict involves pre-registration and posting the parameters of the study in advance at a registry \cite{Sullivan2019-wh, Olevska2021-if, Teo2024-wk}. In contrast, BOP benchmarks give, in principle, a higher degree of neutrality, but the challenge to assess neutrality directly remains. Other benchmarking models are possible, from challenges to hackathons, with the hope that a consensus of the masses can further help to establish neutrality. One potential strategy, in the context of the systematic design of benchmarks, is establishing a system where a benchmark definition is parsed into a pre-registration template. Templates exist for simulation studies that could be adapted more generally for benchmarking projects \cite{Siepe2023-tz}. 

Transparency and clear governance guidelines provide an effective means to detect and react to potential misuse, e.g., by explicit rules to validate results. Other ethical considerations include facilitating safe interactions and fair resource allocation and efficiency. Clear code of conduct statements proactively flagging harassment (e.g., by sex, ethnicity or academic seniority, among others) and provide explicit means to ease collaboration. In terms of efficiency, comparable allocation of resources (e.g., computing power, support) to all contributors, while also being aware of $CO_2$ emissions, could also be spelled out in the governance model.

Mature communities are key to sustaining and reflecting the existing challenges in a given field. For instance, CASP, a continuous benchmark by nature, has a established community that defines the scope of critical assessment, input data, target definitions, and composite and normalized metrics to equally weight multiple performance scores \cite{simpkin2023tertiary}. At the continuous benchmark level, CASP keeps track of past performances and introduces metrics to evaluate possible artifacts, i.e., whether progress during a CASP round could arise from different target difficulty \cite{kryshtafovych2019critical}. To facilitate community contribution, we believe that benchmarking systems can lower the barrier to contributing, hence making the contribution efforts less likely to be biased.

Finally, there is a tension between preregistration and the concept of `continuous' benchmarking, where new datasets, methods and new metrics are added over time. The logic of continuous benchmarking is that as the computational task becomes more understood, it becomes increasingly clear how to best evaluate the performance of methods. 

\section*{Long-term software reproducibility and data preservation}

Software plays an important role in the current scientific landscape \cite{Howison2015-rc}, since it is widely used throughout the scientific process, including data collection, simulation, analysis and reporting. Even if open sourced, however, it is sometimes easier to develop new code than to reuse existing software \cite{Trisovic2022-ol}. This leads to the phenomenon of `academic abandonware', where projects are forgotten in code repositories and not maintained after delivering a research output.

Operational costs of benchmarks are given by the expected compute and storage requirements. In the context of benchmarking, methods under evaluation typically have variable computational demands. Benchmark designers could impose resource limits, thereby evaluating the methods under constraints. Storage costs, on the other hand, are more predictable and can be directly influenced by design decisions of the benchmark (e.g., low retention for re-generatable artifacts, `cold storage' for code archives and software, external data stores such as Zenodo for method performance artifacts, CernVM-FS for reusable software).

\subsection*{Design properties}

A good benchmarking system needs to package its components in a way that ensures that their execution and artifacts can be consistently replicated (within practical limits).

The UNIX design philosophy \cite{Pike1984-we} is worth embracing to produce software for benchmarking systems that are lightweight, robust, and easily maintainable; ease of use and extensibility are also desirable. Furthermore, to reduce the entry barrier for contributors, the benchmarking system should be capable of running locally with minimal setup. This approach ensures that the system remains relevant and functional even in the absence of centralization. Furthermore, intuitive APIs and sensible defaults \cite{Proctor2018-do} are factors that contribute to highly usable systems. Structured approaches to documentation, including frameworks such as diataxis \cite{ProcidaUnknown-vf}, can facilitate adoption and new contributions to a benchmark.

\subsection*{Reproducible software environments \label{softwareenvs}}

Computational reproducibility comes from controlling the triad of data, code, and environment \cite{Hill2024-gf}. During benchmarking, a workflow runs code transforming inputs into outputs; execution is affected by the base system (operating system, compiler toolchains, libraries) and its configuration. The benchmark definition should control the execution environment.

Leaving data aside, it can be useful to divide the codebase used into three distinct categories:

\begin{enumerate}
\item The benchmarking system has an impact on operational costs, such as storage needs, execution platform, etc.  The system mandates the choice of a particular stack and workflow manager. At this level, any requirements for input/output formats and shapes are defined.

\item Benchmark components, like dataset-generating simulations, methods, or metrics, are typically small scripts that process data and use external libraries. Even if method development happens elsewhere, good practices should be embraced (output validation, abnormal termination checks, running on subsets of input data).  Metadata validation should be enforced for each module, including authorship, versioning, and licenses.

\item Software dependencies cause the biggest impact on replicability and maintainability; archiving all combinations of a large dependency tree, across several OS images, exponentially increases retention and maintenance costs. Making the long-term replicability of a benchmark tractable implies adopting a sane software management system.

\end{enumerate}

The responsibility of handling software environments poses an interesting challenge to benchmarking systems. On one hand, some workflow managers provide means to install and activate software and re-run workflows when software specification changes, even if the workflow topology remains the same. For these, software is the workflow manager's responsibility. On the other hand, some benchmarks might be designed to compare the same workflow topology run with different software backends (i.e., with \texttt{conda} vs compiled) or with different versions of a dependency (e.g., \texttt{samtools}/\texttt{htslib} releases). Benchmarking systems should provide enough flexibility to make all variants possible (i.e., decouple the benchmark topology from the software backend).

Finally, scientific software management handling is known for its nuances, including abundant dependencies with known incompatibilities and complex installation recipes \cite{droge2023-ax}. Scientific software management is handled via traditional (Linux) package managers (RPMs, DEBs), automated and reproducible installs (such as \texttt{EasyBuild} \cite{Hoste2012-gg} or \texttt{Spack} \cite{Gamblin2015-ll}), central software stacks and environment modules (i.e., \texttt{Lmod}), \texttt{conda}, containers (e.g., \texttt{apptainer} or \texttt{Singularity}; \cite{Kurtzer2017-mn}), or read-only network filesystems providing ready-to-use installations (\texttt{CernVM-FS}) \cite{Blomer2013-jk}. Further strategies are under active development: EESSI \cite{droge2023-ax} and the Digital Research Alliance of Canada \cite{Boissonneault2019-wm} are effectively providing multi-purpose, reusable and efficient software installations via \texttt{CernVM-FS}; \texttt{conda} is speeding up their deployment procedures (via \texttt{py-rattler}/\texttt{pixi}) and allowing compilation to Web assembly (to execute code within a web browser); and \texttt{apptainer} (former \texttt{Singularity}) provides rootless containerization suitable for running in scientific clusters. 

\section*{Hardware}

Software, both from benchmarking systems and from the benchmarking workflow itself (the benchmarked methods), can be run on different hardware, from commodity computers to dedicated hardware, often large servers or academic and commercial clouds. Performance profiling (e.g., how fast a benchmarked method runs) requires access to fair and comparable resources. 

Comparable resources can be provisioned on demand in cloud computing, or can be ensured by deploying benchmarking systems in controlled and dedicated hardware. Ideally, performance results (e.g., wallclock time) could be bundled to metadata describing the used hardware (e.g., CPU characteristics).

It is worth noting that CPU microarchitecture influences performance, particularly when using optimized compilers to improve performance of benchmarked methods at the expense of making binaries microarchitecture-specific. This feature is not solved by containerization. Benchmarking systems can automate software installations with equivalent degrees of optimization across benchmarked methods, and/or provide multi-architecture builds.

Finally, benchmarking methods could benefit from high performance computing-style user quotas and timeouts to cap the maximum memory and/or CPU cycles each method can use.

\section*{Software licenses and attribution}

Even a simple, single-person benchmark comparing methods relies on multiple contributions, from dataset creators and method developers to system software and workflow authors. Hence, data, code and operating system dependencies are all accessible during the benchmarking process. For reproducibility, transparency and to build trust \cite{Laine2007-py}, this accessibility and freedom should be extended to others to run (re-run), copy (including fork), distribute (e.g., snapshot), study (e.g., read how someone is implementing a method), change (bug fix, contribute) and improve (extend) benchmarks. Such freedoms are granted (or restricted) by software and content licenses, as are copyright and authorship and attribution \cite{Kreutzer2014-ua}. Without explicit licenses, even publicly available code is copyrighted by default, and reuse is restricted (to reproduce, distribute, or create derivative works) \cite{Kreutzer2014-ua}. A concise summary of license rights and restrictions, and compatibility across commonly-used open-source licenses can be found in Table~\ref{tab:licenses}. Benchmark contributions should be licensed and associated to scholarly names including email and ORCID (as the canonical identifier), hence allowing seamless reuse, transparency and attribution. As a consequence, properly-licensed benchmark components (e.g., a procedure to compute a clustering metric; or a clustering method) could be reused across benchmarks; we envision assisted benchmark design tools to leverage and reuse benchmark components, perhaps guided by AI.

\begin{table}[]
    \caption[Overview of commonly-used open-source licenses in benchmarking studies.]{Overview of commonly used open-source licenses in benchmarking studies. Code or data with restrictions to redistribute or modify are not suitable for collaborative benchmarking. \emph{License}: license name. \emph{Redistribution}: right to redistribute, e.g., reuse in another benchmark/somewhere else. \emph{Modification}: right to modify the code to produce derivative works, e.g., in another benchmark/somewhere else. \emph{Sublicensing}: for derivative works, obligation to keep the same license or freedom to be licensed differently (even copyright). Relevant if the code contributor does not want derivative works to be `closed source' or copyrighted. \emph{Commercial use}: for derivative works, whether commercial usage is restricted or not. Relevant to code contributors not willing commercial reuse of their code (by others). \emph{Attribution}: for derivative works, obligation to attribute original author(s). Relevant to code contributors willing their names/authorship to be credited in derivative works. \emph{Linking}: dynamic or static linking / use as libraries. Particularly important for metrics, as these code can be frequently incorporated (=executed) across benchmarks. }
    \vspace{0.5cm}
    \centering
    \begin{adjustbox}{max width=\textwidth}
    \begin{tabular}{|l|l|l|l|l|l|l|}
    \hline
    License       &  Redistribution       &   Modification   & Sublicensing  & Commercial use & Attribution    & Linking   \\ \hline
    no license	& not allowed	& not allowed	& no sublicensing	& not allowed	& required	& not allowed	 \\
    GPLv3 &   	allowed	 & with restrictions	& GPLv3-compatible	& allowed	& required	& GPLv3-compatible \\
    Apache &	allowed &	allowed	& minimal restrictions	& allowed	& required	& allowed \\
    MIT	& allowed	& allowed	& full freedom	& allowed	& required	& allowed	 \\
    Public Domain	& allowed	& allowed	& full freedom	& allowed	& not required	& allowed  \\
    \textsc{CC-BY-NC-ND}	& allowed	& not allowed&	(no derivatives)	& not allowed	& required &	(not a code license)	 \\
    \textsc{CC-BY-SA-NC}	& allowed	& allowed	& not more restrictive& 	not allowed &	required	& (not a code license)	 \\
    \textsc{CC-BY-SA}	& allowed	& allowed	& not more restrictive &	allowed	& required&	(not a code license)  \\
    \textsc{CC-BY} &	allowed	& allowed	& full freedom 	& allowed& 	required&	(not a code license) \\ \hline

    \end{tabular}
    \end{adjustbox}
    \label{tab:licenses}
\end{table}


Similarly, even if all benchmark code were free and open source (FOSS) software, running it could be hindered by (non-free) proprietary interpreters (e.g., MATLAB). Benchmarking systems should be explicit in allowing or disallowing proprietary software, and be aware and track EULAs (end user license agreements). This conundrum extends to the operating system layer and to system dependencies (e.g., compilers). In terms of benchmark re-runnability, fully FOSS system dependencies arguably offer the highest reusability, and can be handled in a reproducible manner across operating systems and system architectures (e.g., amd64, arm64) \cite{droge2023-ax}.

Finally, data files (inputs and outputs) are frequently stored in centralized repositories \cite{Potter2015-fs, Sicilia2017-ko, van-de-Sandt2019-wd} under content licenses, which might enforce attribution or copyleft. A benchmarking system should automate license metadata handling during data import and export, crediting the original authors if applicable; it also represents an opportunity to enforce that academics use appropriate licenses for data and code.

\section*{There are various hidden design tradeoffs}

So far, no single benchmarking system can accommodate all use cases in bioinformatics. However, documentation of design choices for a given system can help to align expectations. Some important design trade-offs are:

\begin{itemize}
    \item Flexibility vs. constraints:  An unconstrained benchmark grants method contributors greater flexibility but requires more effort from the benchmarker (maintenance, version support). Introducing constraints (such as software dependencies or common data formats) adds initial setup work but can boost maintainability through built-in validation and quality checks.
    \item  Achieving software reproducibility needs discipline and commitment \cite{Lamb2021-tb}. A benchmarker can state the desired degree of replicability and constraints. 
    \item  Security concerns: decentralized runs lower the participation barrier, but also increase the attack  surface. Sandboxed environments (restricting network access, filesystem access etc) can be provided as mitigation. Additionally, static and dynamic code analysis in the continuous integration process can aid in approving potentially risky submissions.
    \item Versioning granularity: a single benchmark could address performance of a given method across its versions when run on the same datasets; independently, this benchmark could be updated so several benchmark versions are released. Benchmarking systems should provide versioning at both component and benchmark levels.
\end{itemize}

\section*{Open data formats and standards}

Benchmarking involves applying methods to data. Data types can be diverse and even a single dataset can be stored in multiple formats. A broad distinction can be made according to general vs. specific formats and open vs. proprietary formats. File formats specific to a programming language should be avoided since those formats may change over time, security risks can arise \cite{Huynh2023-pq, Bleih2024-lv}, and they are not always interoperable. Likewise, proprietary formats should be avoided, since they may change and could require specific (proprietary) software to read. General language-agnostic formats often do not change over time and thus align to FAIR principles \cite{Wilkinson2016-bh}. Generally, using established standards is recommended (e.g., SAM, BED, FASTQ, etc for genomics), since related methods are developed around these formats.

If raw input data consists of multiple file formats, an option could be to convert the data into a common format with the caveat that this `intermediate' raw data could be either stored (increased storage usage, easier benchmark definition) or created `on the fly' (increased compute). Furthermore, how raw input data are retrieved (e.g., automatic retrieval once and store locally, automatic retrieval every time, or manual retrieval and store locally) needs to be decided. These intermediate results could also be made FAIR.

Additionally, validation is recommended. Validations include content integrity (e.g., md5sum), adherence to a data specification (e.g., JSON Schema), and data semantics (e.g., SHACL for linked data). 

Finally, metadata (automatic versus manual annotation) should not be neglected. Automatic generation of metadata might require predefined standards and an ontology to fully track how the data was generated \cite{LeboUnknown-om}. 

\section*{Conclusion}
\label{sec:conclusion}

Benchmarking is a cornerstone practice in bioinformatics, impacting how methods are conceived, developed, published and chosen. Benchmarking is driven by the effort of people with the expertise to address biological and technical problems. This expertise can be leveraged by current practices of community benchmarking, so scientific advances are eased by joint efforts to catalog open problems, generate (or simulate) appropriate input data, develop methods to address them, and evaluate their performances. Community benchmarking has been proven successful and  productive \cite{moult2005-ne,Capella-Gutierrez2017-dh,luecken2025open}.

Here, we discussed the motivation and process of community benchmarking at several levels, from definition to execution, while highlighting strategies and technologies to potentially ease the technical and scientific challenges as well as how to organize contributions from communities. 


\section*{Acknowledgments} 

We thank the Robinsonlab members for their constructive feedback.

\section*{Funding}
MDR acknowledges funding from the Swiss National Science Foundation (grants 200021\_212940 and 310030\_204869) as well as support from swissuniversities P5 Phase B funding (project 23-36\_14). CS is supported by the Novartis Research Foundation. The funders had no role in study design, data collection and analysis, decision to publish, or preparation of the manuscript.

\section*{Conflict of interest}

We report no conflict of interest.

\section*{Author contributions}

Conceptualization: all. Figures: IM, BC. Original draft preparation: all. Writing, reviewing and editing: all. Supervision: IM, MDR. Funding acquisition: MDR.



\pagebreak
\linespread{1} 

\bibliographystyle{plos2015}


\end{document}